\documentstyle[prd,aps,amsmath,epsf,graphicx]{revtex}

\DeclareFontFamily{OT1}{rsfs10}{}
\DeclareFontShape{OT1}{rsfs10}{m}{n}{ <-> rsfs10 }{}
\DeclareMathAlphabet{\mathscript}{OT1}{rsfs10}{m}{n}

\newcommand{\be}{\begin{equation}}
\newcommand{\ee}{\end{equation}}

\newcommand{\bea}{\begin{eqnarray}}
\newcommand{\eea}{\end{eqnarray}}
\newcommand{\ba}{\begin{array}}
\newcommand{\ea}{\end{array}}

\newcommand{\ns}{\normalsize}
\newcommand{\pt}{\partial}

\def\a{\alpha}
\def\b{\beta}
\def\g{\gamma}

\def\d{\delta}
\def\e{\epsilon}

\def\f{\phi}

\def\k{\kappa}
\def\l{\lambda}
\def\m{\mu}
\def\n{\nu}

\def\p{\pi}
\def\q{\theta}

\def\r{\rho}

\def\t{\tau}

\def\F{\Phi}

\def\S{\Sigma}

\begin{document}

\begin{titlepage}

\title{
\hfill{\ns SUSX-TH/02-016\\}
\hfill{\ns OUTP-02-16P\\}
\hfill{\ns hep-th/0207281\\[2cm]}
{\Large Five-Dimensional Moving Brane Solutions with Four-Dimensional
Limiting Behaviour}
\\[1cm]}
\setcounter{footnote}{0}
\author{{\ns\large
 Edmund J.~Copeland$^1$\footnote{email: e.j.copeland@sussex.ac.uk}~,
\setcounter{footnote}{1}
 James Gray$^2$\footnote{email: J.A.Gray2@newcastle.ac.uk}~,\\
\setcounter{footnote}{2}
 Andr\'e Lukas$^1$\footnote{email: a.lukas@sussex.ac.uk} and
\setcounter{footnote}{3}
 David Skinner$^3$\footnote{email: skinner@thphys.ox.ac.uk}\\[1cm]}
      {\ns $^1$Centre for Theoretical Physics, University of Sussex}\\
      {\ns Falmer, Brighton BN1 9QJ, United Kingdom}\\[0.2cm]
      {\ns $^2$Department of Physics, University of Newcastle upon Tyne}\\
      {\ns Herschel Building, Newcastle upon Tyne NE1 7RU,
      United Kingdom}\\[0.2cm]
      {\ns $^3$Department of Physics, Theoretical Physics, 
      University of Oxford}\\
      {\ns 1 Keble Road, Oxford OX1 3NP, United Kingdom}}


\maketitle

\vspace{2cm}

\begin{abstract}
Under certain conditions some solutions to five-dimensional heterotic M-theory can be accurately described by the four-dimensional action of the theory - they have a four-dimensional limit. We consider the connection between solutions of four and five-dimensional heterotic M-theory when moving five-branes are present in the bulk. We begin by describing how to raise the known four-dimensional moving brane solutions to obtain approximate solutions to the five-dimensional theory, presenting for the first time the metric template necessary for this procedure. We then present the first solutions to the five-dimensional theory containing moving five-branes. We can then discuss the connection between our new exact five-dimensional solution and the four-dimensional ones. It is shown that our new solution corresponds to a solution with a ${\it static}$ brane in four-dimensions. In other words our new solution could not have been identified as containing a moving brane from a purely four-dimensional view point.
\end{abstract}

\thispagestyle{empty}

\end{titlepage}

\renewcommand{\thefootnote}{\arabic{footnote}}


\section{Introduction}
\label{intro}

Branes moving during the course of the evolution of the universe
constitute an interesting and generic feature of brane-world
models. In this paper, we will analyse this phenomenon in the context
of five-dimensional heterotic brane-world models derived from
heterotic M-theory in eleven
dimensions~\cite{Horava:1996qa,Horava:1996ma}. The five-dimensional
effective action for these brane-world models has been derived
recently~\cite{Lukas:1999hk,Derendinger:2001gy,Brandle:2002ts} and the
associated four-dimensional effective action and its relation to the
five-dimensional theory has been presented in
ref.~\cite{Derendinger:2001gy}. These results open up the possibility
of explicitly studying moving-brane cosmologies in a well-defined
M-theory setting.

In the context of the four-dimensional effective theory, these results
have been applied to obtain the first heterotic moving-brane
cosmological solutions~\cite{Copeland:2001zp}. These solutions can be
viewed as generalisations of the familiar string-theory rolling-radii
solutions (see {\it e.g.}~\cite{Mueller:1990}). In addition to the
dilaton and the universal $T$-modulus they involve the position
modulus of the moving brane. In fact, from the structure of the
four-dimensional effective action, the evolution of the dilaton and
the $T$-modulus cannot be disentangled from the brane motion, or, more
precisely, the dilaton and the $T$-modulus cannot be set to constants
once the brane moves. This has important consequences for the
properties of such moving-brane cosmologies. For example, it was found
in ref.~\cite{Copeland:2001zp} that the effect of the brane is to
interpolate between two rolling-radii solutions with static branes at
early and late time, both in the positive and negative time branch. In
addition, all such four-dimensional solutions become strongly coupled
asymptotically; that is, at early and late time in both branches. From
this result, proposals for an initial state of the universe based on
weak coupling~\cite{Khoury:2001wf} appear to be problematic in the
presence of a moving brane.

In this paper, we will be investigating moving brane cosmological
solutions within the framework of five-dimensional heterotic
M-theory~\cite{Lukas:1999yy,Lukas:1999tt} and the connection to their
four-dimensional counterparts found in ref.~\cite{Copeland:2001zp}.
We begin by showing how to oxidise a moving brane solution of the
four-dimensional theory to an approximate solution of
the five-dimensional theory. Specifically we provide a template for a
five-dimensional metric and dilaton. To obtain the approximate
five-dimensional configuration one simply has to take certain
quantities from the four-dimensional solution and substitute them into
this expression. The resulting solution is accurate to first order in
the slowly moving moduli and weak coupling expansions as will be described.

We then go on to look for five-dimensional moving brane solutions. The
technical challenge here is to find suitable solutions to the
five-dimensional bulk theory in which to embed the branes. For
five-dimensional heterotic M-theory only a very limited number of such
solutions are presently known. The most elementary one is the BPS
domain-wall vacuum of ref.~\cite{Lukas:1999yy}. However, as we have
shown~\cite{Copeland:2001zp}, this solution only allows the embedding
of {\it static} branes, that is, branes which do not move in their
transverse direction. The same is true for the simple separating
cosmological solutions of ref.~\cite{Lukas:1998qs}. We, therefore, use
the only other known class of bulk solutions which were found in
ref.~\cite{Chamblin:1999ya}. We show that, in addition to the
boundaries, a moving brane can be embedded in these bulk solutions.

Finally we investigate the connection between four and
five-dimensional moving brane solutions. We find that our
five-dimensional solutions have a four-dimensional interpretation only
in particular limits. Specifically they can be understood as being
described by the effective four-dimensional theory only for certain
periods of time. Remarkably, while the bulk 5-branes generically move
across the orbifold, the four-dimensional limit requires that they be
static. The 5-brane can remain static for an arbitrarily long period
of time, however if it does start to move the four-dimensional
description breaks down and a brane collision is inevitable. In this
case, the breakdown of the effective description is not caused by
moving to strong-coupling, but rather because the 5-brane
moduli evolves more quickly than is allowed in the lowest-order
four-dimensional action. In other words, studying the five-dimensional
theory has revealed a new type of moving-brane solution which could
not have been identified as such from a purely four-dimensional
viewpoint. Moreover, in the four-dimensional limit the solutions
become more weakly-coupled with time in the positive time branch. This
is in contrast to the  purely four dimensional moving-brane solutions
found in ref.~\cite{Copeland:2001zp} which are always strongly-coupled
asymptotically.

The plan of the paper is as follows. In the next Section, we set the
scene by presenting the minimal version of the five-dimensional
brane-world theory and its associated four-dimensional effective
action.  We also provide the template mentioned above for raising
four-dimensional moving brane solutions to five-dimensions. The
five-dimensional moving-brane solutions will be presented in
Section~\ref{solution}. In Section~\ref{weak}, we analyse their
four-dimensional limits so providing explicit examples illustrating
the transition between four and five dimensions, first without and
then with a bulk 5-brane. We end with a discussion of our results and
our conclusions. Appendix A contains numerical solutions for the
motion of the five-branes and orbifold fixed-points and finally
Appendix B discusses the causal structure of the five-dimensional
solutions.


\section{The Actions in Five and Four Dimensions}
\label{action}

The five-dimensional theory which we will consider emerges as the
low-energy limit of eleven-dimensional Ho\v{r}ava-Witten
theory~\cite{Horava:1996qa,Horava:1996ma}, that is eleven-dimensional
supergravity on the orbifold $S^1/Z_2\times {\mathcal M}_{10}$ coupled
to two ten-dimensional $E_8$ SYM theories each residing on one of the
two orbifold fixed planes. Upon compactification of this theory on a
Calabi-Yau three-fold to
five-dimensions~\cite{Lukas:1999yy,Lukas:1999tt} one obtains a
five-dimensional $N=1$ gauged supergravity theory on the orbifold
$S^1/Z_2\times {\mathcal M}_4$. This bulk theory is coupled to two
$D=4$, $N=1$ theories which originate from the ten-dimensional $E_8$
gauge theories and are localised on 3-branes which coincide with the
now four-dimensional orbifold fixed planes.

In compactifying from eleven to five dimensions, it is possible to
include additional 5-branes in the
vacuum~\cite{Lukas:1999hk,Derendinger:2001gy}. These 5-branes are
transverse to the orbifold and wrap (holomorphic) two-cycles within
the Calabi-Yau three-fold.  Hence, they also appear as 3-branes in the
five-dimensional effective theory. However, unlike the ``boundary''
3-branes which are stuck to the orbifold fixed-points, these 3-branes
are free to move across the orbifold. Altogether, the five-dimensional
theory is then still a gauged $N=1$ supergravity theory on the
orbifold $S^1/Z_2\times {\mathcal M}_4$. It now couples to the two
$N=1$ theories on the boundary 3-branes as well as to a series of
$N=1$ theories residing on the additional 3-branes.

For the purposes of this paper, we will consider the inclusion of only
one intermediate 3-brane. Further, we consistently truncate the action
to the minimal possible set of fields. These are the metric $g_{\a\b}$
and the dilaton $\F$ in the bulk, together with the embedding
co-ordinates of the brane (which, after gauge-fixing, leads to one
physical degree of freedom corresponding to transverse motion of the
brane). We will use co-ordinates $x^\a$, where $\a ,\b ,\dots =
0,\dots ,3,5$ for the five-dimensional space, and allow general
embeddings of the branes within this bulk space. Four-dimensional
co-ordinates $x^\m$ will be indexed by $\m , \n ,\dots = 0,\dots ,3$.

The five-dimensional action contains pieces from the bulk, the two
orbifold fixed-planes (boundaries) and the 5-branes, that is,
\begin{equation}
S_5 = S_{\rm bulk} + \sum_{i=1,2,5} S^{(i)}\; ,
\label{5daction}
\end{equation}
where $i=1,2$ refer to the two boundaries and $i=5$ refers to the
5-brane. Specifically, these pieces are given by~\cite{Brandle:2002ts}
\bea
S_{\rm bulk} &=& -\frac{1}{2\k_5^2} \int_{{\mathcal M}_5}
\sqrt{-g}\left[ \frac{1}{2}R + \frac{1}{4}(\pt \F)^2 + V_{\rm
bulk}(\F) \right] \label{Sbulk}
\\
S^{(i)} &=& -\frac{1}{2\k_5^2}
\int_{\S_i} \sqrt{-h^{(i)}}V_i(\F) \label{Sbrane}. 
\eea
Here, $\k_5$
is the five-dimensional Newton constant and $h^{(i)}$ are the determinants of the four-dimensional
induced metrics on the branes $\S_i$. The branes $\S_1$
and $\S_2$ coincide with the fixed-planes of the orbifold, whilst
$\S_5$ is free to move between these two.

The bulk potential is given by
\begin{equation}
 V_{\rm bulk} = \frac{\a^2}{3} ~ e^{-2\F} \label{V5bulk}
\end{equation}
where $\a$ is a sum of $\q$ functions~\footnote{Ref.~\cite{Brandle:2002ts}  uses
the orbifold picture where the action~\eqref{5daction} and the
definition of $\a$,~\eqref{alpha}, contain an additional term related
to the $Z_2$ mirror of the 5-brane. In this paper, we will be working
in the boundary picture and have, hence, dropped these terms.}
\begin{equation}
 \a = \sum_{i=1,2,5}\a_i\q (\S_i)\; . \label{alpha}
\end{equation}
The boundary potentials $V_i$, where $i=1,2$ take the form
\begin{equation}
V_i = 2\a_i~e^{-\F} \label{Vfp}
\end{equation}
while the potential $V_5$ on the 5-brane reads
\begin{equation}
V_5 = \a_5~e^{-\F}\; . \label{V5branes}
\end{equation}
The constants $\a_i$ can be interpreted as charges on the boundaries
and the 5-brane and they satisfy the cohomology condition
\begin{equation}
 \sum_{i=1,2,5}\a_i = 0\; .  \label{cohom}
\end{equation}

\vspace{0.4cm}

Varying the above action gives the bulk Einstein equation and bulk
scalar field equation in the usual way. The embeddings of the branes
$\S_i$ within this bulk are described by the Israel conditions
\bea
\left[ K_{\a\b}^{(i)}\right] &=& -\frac{1}{3} V_ih_{\a\b}^{(i)}
\label{Israel} \\
\left[ n_\a^{(i)} \pt^\a \F \right] &=& \frac{dV_i}{d\F}.
\label{phiboundary}
\eea 
where $K_{\a\b}^{(i)}$ is the extrinsic curvature and we have defined
$\left[ X \right] \equiv X_+ - X_-$. The jump of any quantity across
the $Z_2$ orbifold-fixed planes is, of course, simply twice the value
next to the boundary, taken with a positive sign at the first boundary
and a negative one at the second.

\vspace{0.4cm}

The vacuum of this five-dimensional theory is a BPS domain-wall
solution first derived in ref.~\cite{Lukas:1999yy}. We may compactify
the five-dimensional theory on this vacuum, promoting certain
constants to moduli, so as to obtain the four-dimensional action of
heterotic M-theory. If one considers only those fields presented in
the action \eqref{Sbulk}-\eqref{Sbrane} (a consistent truncation of the
full five-dimensional theory) then one arrives at the four-dimensional
action~\cite{Derendinger:2001gy}
\begin{equation}
S = -\frac{1}{2\k_4^2}\int\sqrt{-g}\left[\frac{1}{2}R+\frac{1}{4}\pt_\m\f
\pt^\m\f +\frac{3}{4}\pt_\m\b\pt^\m\b+\frac{q_5}{2}e^{\b -\f} \pt_\m
z\pt^\m z\right]\;  \label{4daction}
\end{equation}
where the four-dimensional Newton constant $\k_4$ is related to its
five-dimensional counterpart by
\begin{equation}
\k_4^2 = \frac{\k_5^2}{2\pi\rho}
\label{4dnewton}
\end{equation}
and the constant $q_5= 2\p\r\a_5$ is proportional to the 5-brane
charge.

All three four-dimensional scalar fields $\b, \f$ and $z$ have a
geometrical interpretation in terms of the underlying five- or
eleven-dimensional theory. Concretely, $\p\r e^\b$ specifies the size
of the orbifold, where $\p\r$ is a fixed reference length, while the
volume of the internal Calabi-Yau space is given by $ve^\f$ with a
fixed reference volume $v$. The field $z$ is normalised so it takes
values in the interval $z \in [0,1]$ and, from a five-dimensional
viewpoint, it specifies the position of the three-brane in the
transverse space. Here $z=0$ ($z=1$) corresponds to the first (second)
boundary. At the perturbative level, the fields $\f$, $\b$ and $z$
correspond to flat directions which will typically be lifted by
non-perturbative potentials from various
sources~\cite{Lima:2001jc,Lima:2001nh,Buchbinder:2002ic,Buchbinder:2002pr}. In
this paper, for simplicity, we will not take such non-perturbative
effects into consideration.  Note, that the above action can be
interpreted as the consistent truncation of an $N=1$ supergravity
theory. The K\"{a}hler potential of this theory and its relation to
the component fields has been given in
ref.~\cite{Derendinger:2001gy,Brandle:2002ts}.

The four-dimensional action is correct to first-order in the
strong-coupling expansion parameter, defined as
\begin{equation}
\e \equiv 2\a  \pi \rho e^{\b-\f} \; .
\label{strongcouple}
\end{equation}
This parameter may be interpreted as measuring the strength of
Kaluza-Klein excitations in the orbifold direction, or the strength of
string-loop corrections to the four-dimensional effective action, or
the relative size of the orbifold and the Calabi-Yau space. When $\e
\ll 1$ the theory is in the weakly-coupled regime and the
four-dimensional effective action can provide a good description of
the system. However, if $\e \sim 1$ or greater, then the
four-dimensional action~\eqref{4daction} cannot be expected to provide
a good description of the theory, as excitations of the
five-dimensional Kaluza-Klein modes become important. The
five-dimensional action~\eqref{5daction}-\eqref{Sbrane} should also
receive higher order corrections in $\e$ governed by the
eleven-dimensional theory. However, it can be expected to provide a
more accurate description of the system at $\e >1$ than the
four-dimensional theory. In this sense, we will use the
action~\eqref{5daction} to obtain a qualitative picture for
brane-evolution at $\e >1$.

Notice also that from a four-dimensional perspective, a moving
5-brane necessarily implies time-evolution of the fields $\f$ and
$\b$, as can be seen from the kinetic term of the $z$ field
in~\eqref{4daction}. The complete set of cosmological solutions for
the action~\eqref{4daction} (assuming Ricci flat spatial sections) has
been given in \cite{Copeland:2001zp}. It has been shown for those
solutions, that a moving brane necessarily implies strong coupling,
{\it i.e.} diverging $\e$, asymptotically at both  early and late time. This
property is a direct consequence of the coupling between $z$, $\f$ and
$\b$ in~\eqref{4daction}.

\vspace{0.4cm}

How does one raise a solution of this four-dimensional action to
obtain an approximate solution to the five-dimensional one? As is
well-known, the vacuum solution contains non-trivial dependence on the
orbifold co-ordinate due to the presence of the branes. This
dependence gets modified slightly in the presence of a moving
5-brane. To raise a four-dimensional moving brane solution to
obtain a five-dimensional approximate solution, one must use the
following template. Using co-ordinates in which the orbifold fixed-planes
lie along lines of constant $y$ (at $y=0$ and $y=\p\r$), the
template is
\bea
\label{BPSmet}
ds^2 &=& e^{- \b} g_{\mu \nu} \left(1+ \e B(y) \right) dx^{\mu}dx^{\nu}
+ e^{2 \b} \left(1 +  \e B(y) \right)^4 dy^2
\\
e^{\F} &=& e^{\f} \left(1 + \e B(y) \right)^3
\label{BPSf}
\eea
where, for a 5-brane located at $y = Y = z \p\r$, the function $B(y)$
is given by
\begin{equation}
B(y) = \begin{cases}  \frac{e^{3 \b} \a_5 \dot z^2}{6 \a_1} \frac{y^2}{2}
-\frac{1}{3 \pi \rho}y + c_1 \hspace{4cm}  0 \leq y \leq Y  
\cr 
-\frac{e^{3 \b} \a_5 \dot z^2}{6 \a_2} \frac{y^2}{2} - \left(
\frac{1}{3\pi \rho} - \frac{\pi \rho e^{3 \b} \a_5 \dot z^2}{6 \a_2}
\right)y + c_2 \hspace{1.4cm} Y \leq y \leq \pi\rho
\end{cases}
\label{hdef}
\end{equation}
where
\begin{eqnarray}
c_1 &=& \frac{1}{6} \left(1 + \frac{\a_5}{\a_1}\left(1-z\right)^2
+\frac{e^{3 \b} \a_5}{2 \a_1} (\p\r \dot z)^2
\left(\frac{2}{3} + z^2 - 2 z \right) \right)
\\
c_2 &=& -\frac{1}{6} \left(-1+ \frac{\a_5}{\a_2} z^2
 + \frac{e^{3\b} \a_5}{2 \a_2} (\p\r \dot z)^2 
\left( \frac{2}{3} + z^2 \right) \right) \; .
\label{adef}
\end{eqnarray}
Notice that the five-brane may effectively be removed from this
template by setting $\a_5=0$; for constant moduli we then recover the
standard BPS vacuum solution of~\cite{Lukas:1999yy}. A BPS
configuration is also obtained if the five-brane is static, so $\dot z
= 0$.

\vspace{0.4cm}

To obtain five-dimensional solutions with dynamical five-branes one
simply has to substitute the fields $g_{\m\n}$, $\b,\f$ and $z$ from
the four-dimensional moving brane solutions of~\cite{Copeland:2001zp}
into this template. The resulting configuration will be a solution to
the five-dimensional theory to first order in the weak coupling
and slowly moving moduli approximations.
  

\section{A Five-Dimensional Solution with a Moving 5-brane}
\label{solution}

To date, it has proven elusive to show explicitly the connection
between solutions of the theories~\eqref{5daction}--\eqref{Sbrane}
and~\eqref{4daction} for anything other than the BPS vacuum
state and the separating cosmological solutions of~\cite{Lukas:1998qs}
(the four-dimensional solutions corresponding to which are marked in
figure~\ref{fig1}). Unsurprisingly, the main reason for this is that complete
five-dimensional solutions are extremely difficult to find. One
approach that has been followed by Chamblin \& Reall in
ref.~\cite{Chamblin:1999ya} is to consider bulk spacetimes that only
depend on a single co-ordinate. As noticed in
ref~\cite{Langlois:2001dy}, the constraint that the bulk fields only
depend on one co-ordinate is equivalent to adoption of the ansatz
\begin{equation}
e^{\F} \propto R^6. \label{constraint}
\end{equation}
relating the dilaton $\F$ and brane scale-factor $R$, where the sixth
power is governed by the particular form of the universal potentials
in heterotic M-theory~\eqref{V5bulk}-\eqref{V5branes}. The first steps
towards more general solutions have been taken in
ref~\cite{Charmousis:2001nq}. Although we will not attempt to use such
more complicated solutions explicitly in this paper, in the final
section we will point out how we anticipate they play a r\^{o}le.

\vspace{0.4cm}

Instead, here we wish to generalize the solution of
ref.\cite{Chamblin:1999ya} so as to describe the possibility of
additional 5-branes which are free to move across the orbifold. For
simplicity, we will consider the inclusion of only one bulk 5-brane,
the generalization being straightforward. The metric that achieves
this is given by
\begin{equation}
ds^2 = -\frac{1}{U_\pm(t_\pm)} dt_\pm^2 + U_\pm(t_\pm)dr_\pm^2 +
R_\pm (t_\pm)^2 d{\bf x}^2
\label{5dmetric}
\end{equation}
where co-ordinates $(t_-,r_-)$ and $(t_+,r_+)$ are valid on the left
and right of the 5-brane, respectively. The dilaton field has a
corresponding solution
\begin{equation}
\F = \F_{0\pm} + \frac{6}{7}\ln(t_\pm)
\label{5dphi}
\end{equation}
where $\F_{0\pm}$ are arbitrary integration constants. The functions
$U_\pm$ and $R_\pm$ that appear in the metric are given by
\bea
U_\pm(t_\pm) &=&
49t_\pm^{\frac{2}{7}} \left( 2M_\pm t_\pm^{\frac{2}{7}} - \frac{\a_\pm^2
e^{-2\F_{0\pm}}}{9}\right)  \label{Udef}
\\
R_\pm(t_\pm) &=& t_\pm^{\frac{1}{7}} \; .
\label{Tdef}
\eea 
In this paper, we will mainly consider the region above the
horizons where
\begin{equation}
t_\pm > t_{\pm}^h \equiv \left(\frac{\a_\pm^2
e^{-2\F_{0\pm}}}{18M_\pm}\right)^{\frac{7}{2}}
\label{t_hdef}
\end{equation}
in each portion of the bulk and hence the $U(t)$'s are positive,
making $t$ and $r$ locally timelike and spacelike, respectively. The
conventions in ref.~\cite{Chamblin:1999ya} were the opposite. The
parameters $M_\pm$ are positive, but otherwise arbitrary constants of
integration which may be interpreted as the mass of the singularity at
$t=0$ in the maximally extended spacetimes. They may be chosen
independently on each side of a bulk 5-brane. The `charges' felt in
each region of the bulk are given by $\a_- = \a_1$ and $\a_+ = \a_1 +
\a_5 = -\a_2$ in terms of the charges on the branes. This solution is
essentially two copies of the spacetime found in
ref.~\cite{Chamblin:1999ya}, sewn together at the bulk brane
$\S_5$. The sewing is continuous, but not necessarily smooth as the
5-brane supports energy-momentum through its potential. Continuity of
the metric requires that $t_-(\S_5)=t_+(\S_5)$ (we can make this
choice of scaling of ${\bf x}_{\pm}$ without loss of
generality). Continuity of the dilaton then forces us to choose
$\F_{0-} = \F_{0+}$. In the limit $M_\pm \rightarrow 0$ this solution
becomes the BPS vacuum of ref.~\cite{Brandle:2002ts,Lukas:1999yy}.

\vspace{0.4cm}

Since $R_\pm$ are monotonically increasing functions of $t_\pm$ alone,
the branes may be taken to be located in the bulk such that their
scale factors are $R=R_1$, $R_5$ and $R_2$ respectively, with
$R_1<R_5<R_2$. The locations of the branes are governed by the Israel
conditions~\eqref{Israel}. Using the co-ordinates of the
metric~\eqref{5dmetric} it is possible to write these in a convenient
form. Following the procedure of ref.~\cite{Chamblin:1999ea} where
non-BPS brane configurations in type IIA massive supergravity were
considered, we twice square the $ij$ components of eq.~\eqref{Israel} to
obtain the brane equations of motion as
\begin{equation}
\left(\frac{dR_i}{d\l}\right)^2 = \frac{9e^{2\F_0}[M]^2}{[\a]^2}
\frac{1}{R_i^6} + \left(\{M\} -
\frac{\{\a\}[M]}{[\a]}\right)\frac{1}{R_i^8}
\label{breom}
\end{equation}
where $\l$ is proper time measured on the relevant brane.
We have defined $\left\{ X \right\} \equiv X_+ + X_-$, and $[\a] = \a_i$
is the (bulk 5-) brane's charge.

Notice that for a brane at an orbifold fixed-plane, $Z_2$-symmetry
requires that $[M] = 0$ and $\{M\} = 2M$, whereas for a bulk
five-brane neither condition is necessarily true. This crucial
difference between the embeddings of the orbifold fixed-planes and
bulk 5-brane immediately leads to the conclusion that, if $M_\pm$ are
chosen to be different in the two sections of the bulk, the 5-brane
will always collide with a fixed plane in at least one asymptotic
regime. This may be seen from the first term in eq.~\eqref{breom}
which dominates as $R_i \rightarrow \infty$. Since this term can only
exist for a bulk brane, the 5-brane will move across the orbifold and
ultimately collide with one of the fixed planes. Conversely, the only
possibility for obtaining a solution where the 5-brane is
asymptotically static with respect to the orbifold is to choose
$M_-=M_+$. Then, as for the fixed planes, eq.~\eqref{breom} may be
solved in compact form and it is clear that the brane separations
approach constants as $R \rightarrow \infty$. In this case the branes'
trajectories in either $(r,t)$ plane are given by
\begin{equation}
r = r_0^{(i)} \pm \frac{(t^h){\frac{3}{7}}}{28 M} \left[ \left(
\frac{t}{t^h}\right)^{\frac{2}{7}} + \ln \left( \left(
\frac{t}{t^h}\right)^{\frac{2}{7}} - 1 \right)\right]
\label{traj}
\end{equation}
where $r_0^{(i)}$ is an arbitrary integration constant and $t^h$ was
defined in eq.~\eqref{t_hdef} using either the $(r_-,t_-)$ or
$(r_+,t_+)$ co-ordinates.

We shall make use only of certain limits of the fixed plane and
5-brane motion in the next section, for which simple analytical forms
can be found in both cases. Diagrams illustrating the orbifold
fixed plane and 5-brane trajectories for more general choices of
parameters may be found in Appendix A.


\section{The four-dimensional limit}
\label{weak}

In this Section we will show how the solution
\eqref{5dmetric}-\eqref{5dphi} is related to solutions of the
four-dimensional action~\eqref{4daction}. First, we will consider the
case without a bulk brane. This provides both the background in which
we later wish to study the 5-brane motion, and also in itself
demonstrates an example of a correspondence between a five- and
four-dimensional solution in a concrete manner. We shall then proceed
to include a five-brane in the bulk.

\subsection{The case without bulk 5-branes}
\label{background}

\begin{figure}[ht]\centering
\includegraphics[height=9cm,width=11cm, angle=0]{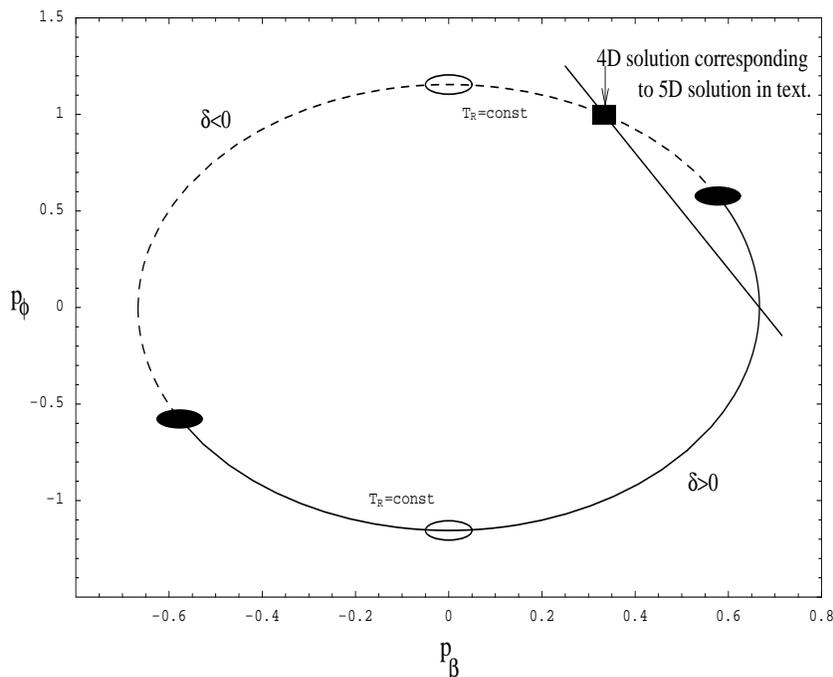}
\caption{\emph{The ellipse of possible expansion powers
$(p_{\b},p_{\f})$ is shown. Solutions on the dotted part of the
ellipse evolve towards weak coupling as $|\t-\t_0| \rightarrow \infty$
whereas those on the solid curve tend towards strong coupling. The
straight line indicates by its intersection with the ellipse the two
four-dimensional solutions which are compatible with the relationship
between $R$ and $\F$ in the five-dimensional solution as given
by~\eqref{constraint}. The four-dimensional solution to which the
configuration given in section \ref{solution} corresponds is marked
with a rectangle. The two points marked on the ellipse with solid black oblongs are the rolling radii solutions which correspond to the separating solutions mentioned at the start of section~\ref{solution}.}}
\label{fig1}
\end{figure}

Homogeneous, isotropic solutions of the four-dimensional
action~\eqref{4daction} for $z=$ const have been extensively studied
(see {\it e.g.}~\cite{Mueller:1990,Copeland:2001zp,Brandle:2000qp}).
Here we will be interested in those which possess metrics with spatially
flat sections, described by
\begin{equation}
\label{4dmetric}
ds^2 = -d\t^2 + e^{2A}d{\bf x}^2
\end{equation}
with the moduli fields
\bea A(\t) &=& A_0 + \frac{1}{3} \ln
\left|\frac{\t-\t_0}{T} \right| \label{Adef} \\
\b(\t) &=& \b_0 + p_{\b}
\ln \left|\frac{\t-\t_0}{T} \right| \label{bdef} \\
\f(\t) &=& \f_0 +
p_{\f} \ln \left|\frac{\t-\t_0}{T} \right|
\label{fdef}
\eea
where $A_0$, $\b_0$ $\f_0$, $\t_0$ and $T$ are arbitrary constants
(clearly they are not all physical, but we shall find it convenient to
express the four-dimensional metric in this form).  On the other hand,
the constants $p_\b$ and $p_\f$ satisfy the constraint $3p_{\b}^2 +
p_{\f}^2 = 4/3$, so the space of possible four-dimensional
rolling-radii solutions may be described by an ellipse as in
figure~\ref{fig1}. The solutions are classified by $\d \equiv p_\b -
p_\f$, with the $\d>0$ solutions evolving towards strong coupling as
$|\t-\t_0| \rightarrow \infty$ and the $\d<0$ solutions becoming
strongly coupled as $|\t-\t_0| \rightarrow 0$. To first order in the
strong-coupling parameter and the slowly evolving moduli expansions,
these solutions may be oxidised to provide solutions to the full
five-dimensional theory by considering them as fluctuations around the
BPS vacuum. To do this one can use a simpler version of the
template~\eqref{BPSmet}-\eqref{BPSf} where the bulk five-brane has
been removed - by setting $\a_5=0$ for example. We now wish to
demonstrate that, in a certain regime, the
solution~\eqref{5dmetric}-\eqref{5dphi} (in the absence of a
five-brane) is equivalent to an oxidised four-dimensional solution.

\vspace{0.4cm}

The question immediately arises as to which of the four-dimensional
solutions we should oxidise. The relationship between $R$ and $\F$
given in~\eqref{constraint} scales to become the constraint
\begin{equation}
p_\f = 2 - 3p_\b
\label{f6ascaled}
\end{equation}
in four dimensions. Therefore, if the five-dimensional fields of
Section~\ref{solution} are to evolve into any of the four-dimensional
rolling radii solutions, then it must be one of the two intersection
points in figure~\ref{fig1}. Which of these two points is the
relevant one can be determined by
noticing that, as $t \rightarrow \infty$, the terms in the
metric~\eqref{5dmetric} that involve the brane charge $\a$ become
negligible, so the solution must become weakly coupled at late
time. This uniquely determines that the required four-dimensional
moduli fields have expansion powers $p_\f = 1$ and $p_\b = 1/3$.

\vspace{0.4cm}

Our procedure shall now be to insert
eqs.~\eqref{4dmetric}-\eqref{fdef} into
eqs.~\eqref{BPSmet}-\eqref{BPSf}, in the absence of a five-brane, to
obtain the oxidised four-dimensional solution. This we will compare
with the full solution of Section~\ref{solution} (for the case where
there is no bulk five-brane of course). We do not expect that the
oxidised expressions will reproduce the entire five-dimensional
solution, but only a certain region thereof. More specifically, since
the oxidised solution is accurate to linear order in the
strong-coupling expansion parameter $\epsilon$, and we have seen that
this limit of the bulk metric occurs at `late' times, it is clear that
the appropriate region of the metric~\eqref{5dmetric} with which to
compare the raised expressions is where $t \gg t^h_{\pm}$. Then each
orbifold fixed-plane is far `above' the horizon that would be present in the
maximally extended bulk.

In the absence of a bulk brane, to first order in the strong coupling
parameter $\e$, the oxidised dilaton has the form
\begin{equation}
e^{\F} = e^\f \left(1 + 2he^{\b-\f}\right)
\label{oxf}
\end{equation}
where $h$ has the simple form $h(y) = \a_1 (|y| - \p\r /2)$ as
appropriate for the BPS vacuum without 5-branes. Inserting
equations~\eqref{bdef} and~\eqref{fdef} into this and comparing with
the five-dimensional expression eq.~\eqref{5dphi} shows that we must
choose the $t$ co-ordinate of equation~\eqref{5dmetric} to be given in
terms of rolling radii co-ordinates employed in
equations~\eqref{BPSmet}-\eqref{BPSf} by
\begin{equation}
t = \left|\frac{\t-\t_0}{T} \right|^{\frac{7}{6}} \left(1 +
\frac{7}{3}h e^{\b_0-\f_0} \left|\frac{\t-\t_0}{T}\right|^{-\frac{2}{3}}
\right)
\label{trans1}
\end{equation}
to first order in brane charges, and we have set $\F_0 = \f_0$.

We recall that the reduction of the five-dimensional
action~\eqref{Sbulk}-\eqref{Sbrane} to the four-dimensional effective
action~\eqref{4daction} was done using co-ordinates where the orbifold
fixed-planes lay at constant $y$, whereas in Section~\ref{solution}
the fixed-planes followed the trajectories given in
eq.~\eqref{traj}. This determines the other change of variable required as
as
\begin{equation}
dr = \g \, dy \pm \frac{2}{7}t^{-\frac{5}{7}}
\frac{t_h^{\frac{1}{7}}}{28M} \, dt
\label{trans2}
\end{equation}
where $\g$ is a possible scaling of the $y$-axis to be determined
later, and again we have only considered terms up to first order in the
strong-coupling parameter.

We can now insert these co-ordinate transformations into the limit of
the five-dimensional solution described above to obtain
\begin{multline}
ds^2 = e^{-\b_0}\left|\frac{\t-\t_0}{T}\right|^{-\frac{1}{3}} \left(1+
\frac{2h}{3} e^{\b_0-\f_0}
\left|\frac{\t-\t_0}{T}\right|^{-\frac{2}{3}}\right) \left[ -d\t^2 +
e^{2A_0} \left|\frac{\t-\t_0}{T}\right|^{\frac{2}{3}} d{\bf x}^2
\right] \\
+ e^{2\b_0}\left|\frac{\t-\t_0}{T}\right|^{\frac{2}{3}} \left( 1 +
\frac{4h}{3} e^{\b_0-\f_0}
\left|\frac{\t-\t_0}{T}\right|^{-\frac{2}{3}}\right) dy^2
\label{limit1}
\end{multline}
for the metric and
\begin{equation}
e^{\F} = e^{\f_0}\left|\frac{\t-\t_0}{T}\right|\left(1+ 2h
e^{\b_0-\f_0} \left|\frac{\t-\t_0}{T}\right|^{-\frac{2}{3}}\right)
\label{limit2}
\end{equation}
for the dilaton. In deriving these expressions, we have made the
following identifications between five- and four-dimensional constants
\bea
\label{iden1}
2M &=& \frac{e^{\b_0}}{(6T)^2}
\\
\label{scale}
\g \; \; &=& \frac{\pm e^{\b_0}}{7\sqrt{2M}}
\\
\label{iden2}
2A_0 &=& \b_0
\eea
The last of these conditions is simply a consequence of a choice of
scaling we have made for the $x_i $ (this choice was mentioned briefly
in the paragraph underneath~\eqref{t_hdef}).

We now see that the late time limit of our five-dimensional solution
(as given in eqs~\eqref{limit1}-\eqref{limit2}) is, to first order in
the strong-coupling parameter, exactly the same as the oxidised
four-dimensional solution (obtained by
inserting~\eqref{4dmetric}-\eqref{fdef} into the template of
eqs~\eqref{BPSmet}-\eqref{BPSf} with $\a_5 = 0$ as appropriate if no
5-brane is present) except for the co-efficient of $dy^2$. In
eq.~\eqref{limit1} this co-efficient contains a factor of
$\frac{4}{3}$ whereas the template has $\frac{8}{3}$. This difference
is to be expected in general when comparing such solutions for the
following reason. If we insert the template structure into the
five-dimensional Friedmann equation and expand to first order in
strong-coupling we obtain,
\begin{equation}
\left( \dot{A}^2 - \frac{\dot{\b}^2}{4} - \frac{\dot{\f}^2}{12}
\right) + \frac{1}{3}\left( 6\dot{A} - 2\dot{\b} -
\dot{\f}\right)\left(\dot{\b} - \dot{\f}\right) h e^{\b-\f} =
\frac{1}{3}e^{-2\b-\f} h'' \mp \sum_{i=1}^{2} \frac{\a_i}{3} e^{-2\b-\f}
\d(y-y_i) \; .
\label{5dFried1}
\end{equation}
The only terms in this
equation which depend on the precise factor in the co-efficient of
$dy^2$ in the metric are,
\begin{equation}
\frac{1}{3}\left( 6\dot{A} - 2\dot{\b} -
\dot{\f}\right)\left(\dot{\b} - \dot{\f}\right) h e^{\b-\f}
\label{nocancel}
\end{equation}
Notice however that these terms are first order both in the
strong-coupling parameter and in squared time derivatives of
moduli. The four-dimensional action~\eqref{4daction} is only accurate
to first order in each of these expansions and so these terms are
negligible when it is valid. Thus in connecting the two solutions the
value of this factor in the metric is irrelevant.

\vspace{0.4cm}

Hence we have established the connection between the two
solutions. The late time behaviour of the metric~\eqref{5dmetric} and
dilaton~\eqref{5dphi} is described by the four-dimensional rolling
radius solution of eqs.~\eqref{4dmetric}-\eqref{fdef} with $p_\f = 1$
and $p_\b = 1/3$. Conversely, one full continuation of this particular
weakly coupled solution into the strong-coupling regime is given
by~\eqref{5dmetric}-\eqref{Udef}. To our knowledge, this is the first
time such a non-trivial correspondence between dynamical solutions of
five- and four-dimensional heterotic M-theory has been explicitly
demonstrated.

\subsection{Including a 5-brane}
\label{brane}

We may now proceed to include a bulk 5-brane in this process of
identification. Four-dimensional solutions of the
action~\eqref{4daction} including the modulus $z$ describing the
location of the 5-brane were studied in~\cite{Copeland:2001zp}. There,
it was shown that $z$ evolves according to
\begin{equation}
z(\t) = z_0 + d \left( 1 + \left| \frac{\t-\t_0}{T} \right|^{p^i_\b - p^i_\f}
\right)^{-1}
\label{zmotion}
\end{equation}
with $z_0$ and $d$ arbitrary constants. $p^i_{\b}$ and $p^i_{\f}$ are
the expansion powers of the  T modulus and the dilaton at infinity. In
our case we know that as we let $t \rightarrow \infty$ we also have
$\e \rightarrow 0$. This is impossible for a four-dimensional solution
which contains a moving brane and so we set $d=0$. Therefore we will
see that, although the 5-brane moves in the full five-dimensional
solution, it corresponds to a solution which has a stationary 5-brane
in the four-dimensional limit. In this limit the expansion powers will
be the same as for the case without a five brane, {\it i.e.} $p_{\phi}
= 1$, $p_{\b}= \frac{1}{3}$.

\vspace{0.4cm}

From the five-dimensional perspective, there is no $Z_2$ symmetry
across the bulk brane and so {\it a priori} we should consider all the
terms in the Israel conditions:
\begin{equation}
\left(\frac{dR_i}{d\l}\right)^2 = \frac{9e^{2\F_0}[M]^2}{[\a]^2}
\frac{1}{R_i^6} + \left(\{M\} -
\frac{\{\a\}[M]}{[\a]}\right)\frac{1}{R_i^8}
\label{breom2}
\end{equation}
where, again, $\l$ is proper time as measured on the relevant brane.
If we make the {\it choice} $\left[ M \right] = 0$, as we are free to
decide, then the 5-brane would move in the maximally extended bulk in
precisely the same way as an orbifold fixed-plane. It is then clear
that as $t \rightarrow \infty$ this solution has a four-dimensional
limit exactly as before, except of course that the function $h(y)$ in
eqs.~\eqref{limit1}-\eqref{limit2} should now be replaced by the
expression given in eq.~\eqref{hdef} as appropriate for raising a
four-dimensional solution on a background containing a 5-brane (with
$\dot z = 0$). In the four-dimensional regime, this corresponds to
choosing $ \pi \rho z_0 = Y$ in eq~\eqref{zmotion}.

On the other hand, for any $\left[ M \right] \neq 0$, it is inevitable
that the 5-brane starts to move with respect to the orbifold
fixed-planes, ultimately colliding with one of them as discussed
above. This effect becomes important after the time
\begin{equation}
t^{\frac{2}{7}} \sim t^{\frac{2}{7}}_{{\rm move}} \equiv
\frac{[ \a ]}{9e^{2\F_0}[M]^2} \left( \{M\}[\a] - \{a\}[M] \right)
\label{tmove}
\end{equation}
for a 5-brane of charge $\left[ \a \right] = \a_5$. This time may be
delayed arbitrarily by taking $\left[ M \right] \rightarrow 0$. In
particular, it is possible to choose $\left[ M \right]$ such that
$t_{\rm move} \gg t^h$ so that the whole model can have evolved well
into the weakly coupled regime long before the 5-brane motion becomes
significant. Of course, for the case where $\left[ M \right]$ is large
enough that $t^h > t_{\rm move}$ the bulk brane will perform its
motion within the five-dimensional regime to start off with, as even
the bulk fields have not evolved sufficiently to allow a four-dimensional
description.

Assuming that $\left[ M\right]$ is sufficiently small, when the
5-brane does begin to move appreciably we may describe it in terms of
the same four-dimensional co-ordinates as were introduced earlier in
eqs.~\eqref{trans1}-\eqref{trans2}. Our above analysis tells us that
this motion should not correspond to that of a four dimensional
solution and we shall now see that this is indeed the case. Writing
the five brane motion in these coordinates we obtain,
\begin{equation}
\frac{dz}{d\t} = \pm \frac{e^{-\frac{3\b_0}{2}}}{\p \r} \left|
\frac{\t-\t_0}{T}
\right|^{-\frac{1}{2}}
\label{5motion}
\end{equation}
where we have considered only the lowest-order terms in the
strong-coupling parameter.  Comparing this expression with
eq~\eqref{zmotion} as found directly from solutions of the
four-dimensional action shows that the 5-brane motion considered here
is not a solution of the weakly-coupled action.

Thus the brane's motion is a higher-order effect than is contained in
the action~\eqref{4daction} and so when such motion becomes
significant our five-dimensional solution is no longer in a
four-dimensional limit due to a break down of the slowly moving moduli
approximation.  To see this in more detail, let us again consider the
five-dimensional Friedmann equation with the raising template metric,
this time including an additional source for the 5-brane. We find
\begin{multline}
3 e^{\b}\left(\dot{A}^2 - \frac{\dot{\b^2}}{4} -\frac{\dot{\f}^2}{12} \right)
- \frac{3}{2} e^{-2\b} \e B'' =
\sum_{i=1}^2  2\a_i e^{-\b-\f} \delta \left( y - y_i
\right) + e^{-\b - \f} \delta \left( y - \pi \rho z \right)  \a_5 
\left(1+ e^{3 \b} \dot{z}^2 (\pi \rho)^2 \right) 
\label{5dFried2}
\end{multline}	
where we keep only terms up to first order in the
strong-coupling expansion parameter and slowly moving moduli approximations.

The zeroth-order contributions survive to appear in the
four-dimensional theory. The terms
\begin{equation}
\frac{3}{2} e^{-2\b} \e B'' +
\sum_{i=1}^2  2\a_i e^{-\b-\f} \delta \left( y - y_i
\right) + e^{-\b - \f} \delta \left( y - \pi \rho z \right)  \a_5 
\left(1+ e^{3 \b} \dot{z}^2 (\pi \rho)^2 \right)
\label{cancel}
\end{equation}
cancel exactly because of the complicated structure of $B(y)$
in the metric template in the presence of a bulk 5-brane (see
eqs.~\eqref{hdef}-\eqref{adef}). However, when the 5-brane starts to
move, we see that there is no term which can match the contribution of
its kinetic energy. Furthermore, since
\begin{equation}
e^{3\b} \dot{z}^2 = 1
\label{expansion}
\end{equation}
when $t  > t^h$ and $t > t_{\rm move}$, the contribution of terms we have
ignored in working to first order in the expansion in the time
derivatives of the $z$ modulus become just as important as those which
we have kept. Thus we see that higher-order corrections to the
template structure would indeed be needed to describe the solution as
soon as the brane motion is important. Hence, as $\t \rightarrow
\infty$, the motion of the bulk brane in this particular
five-dimensional solution does not admit a description in terms of the
lowest order four-dimensional theory. Notice incidentally that
choosing the jump of $M$ to be small only delays the time when the
bulk brane starts to move, and does not `slow it down' at this level
of approximation.

\vspace{0.4cm}

In conclusion, we have established that in a certain regime (namely
when $t^h \ll t \ll t_{\rm move}$) our five-dimensional moving brane
solution corresponds to an oxidised four dimensional rolling radius
solution with a static 5-brane. To our knowledge this is the first
non-trivial example where such a correspondence has been established
explicitly, with or without a bulk brane. If $M_1 = M_2$, the bulk
5-brane remains static for all times, but any mismatch of these
parameters will ultimately always cause the brane to move, forcing us
to consider higher-order terms than are usually kept in the
four-dimensional theory. The matching procedure presented in this
section can be simply extended to a solution with any number of
5-branes present in the bulk.


\section{Discussion \& Conclusions}

We have provided three main new pieces of information in this paper.
Firstly, we have described how to raise four-dimensional moving brane
solutions of heterotic M-theory to obtain five-dimensional solutions
accurate to first order in the $\e$ and slowly moving moduli
expansions. The procedure for raising such solutions is to take
certain quantities such as the values of various fields in the
four-dimensional description and substitute them into the template of
a five-dimensional configuration which we have provided. Secondly, we
have presented the first solutions to five-dimensional heterotic
M-theory which containing moving five-branes.  These are exact
solutions to the full five-dimensional action of the theory.  Finally,
we have shown that at late times on the positive time branch, our
solutions possesses a mild enough warping to admit a four-dimensional
effective description. However, the presence of a 5-brane in the bulk
can destroy this four-dimensional limit as its position modulus can
vary rapidly in the full solution, breaking the slowly moving
moduli approximation necessary for such a description to be valid. We
end up with our five-dimensional solutions generically having a
four-dimensional limit for some intermediate period of time - where
$\e$ has become small enough to admit a weakly coupled description but
the five-brane has yet to start moving significantly. The fact that a
five-dimensional moving five-brane solution corresponds to a
four-dimensional solution with a constant $z$ modulus is
interesting. It means that we could not have identified our solutions
as containing a moving five-brane from a four-dimensional perspective
- the full five-dimensional theory is required for that.
\vspace{0.4cm}

It is of interest to consider briefly how we might expect other
five-dimensional solutions to behave. As pointed out by Chamblin \&
Reall in ref.~\cite{Chamblin:1999ya}, there is of course no reason to
expect that a general solution to Einstein's equations should only
depend on one bulk co-ordinate, even if it is homogeneous and
isotropic in directions parallel to the orbifold planes. More
recently, in ref.~\cite{Charmousis:2001nq} Charmousis has attempted to
construct heterotic M-theory solutions where the bulk fields cannot,
in any gauge, be written as functions of only one co-ordinate, even
locally. Of course, not all of these solutions will necessarily
possess four-dimensional descriptions in any regime, however from the
perspective of this paper we expect that at least some should display
similar behaviour to the $e^{\F} = R^6$ solution considered
here. Specifically, as either $|t| \rightarrow \infty$ or $t
\rightarrow 0$ they should evolve to become one of the other
four-dimensional rolling radii solutions displayed in
fig.~\ref{fig1}. If these solutions were augmented to include 5-branes
across the orbifold, then again we would expect there to be additional
conditions on the validity of the four-dimensional description
dependent upon the velocity of the 5-brane. In particular, it should
in principle be possible to find the full, non-linear five-dimensional
lift of the moving brane solutions in
ref.~\cite{Copeland:2001zp}. Unfortunately, owing to the difficulty of
the five-dimensional field equations, at present we are unaware of any
explicit examples of such solutions.


\section{ACKNOWLEDGMENTS}

J.G. would like to thank Ian Moss for a useful conversation about the
causal structure. D.S. is supported by a PPARC studentship, J.G. is
supported by a Sir James Knott fellowship and A.L. is supported by a
PPARC Advanced Fellowship. D.S. would like to thank E.J.C, A.L. and
Jocelyn Retter for hospitality at Sussex while part of this work was
being completed.

\appendix

\section{Numerical Solutions for brane motion}

\begin{figure}[ht]\centering
\includegraphics[height=8cm,width=11cm, angle=0]{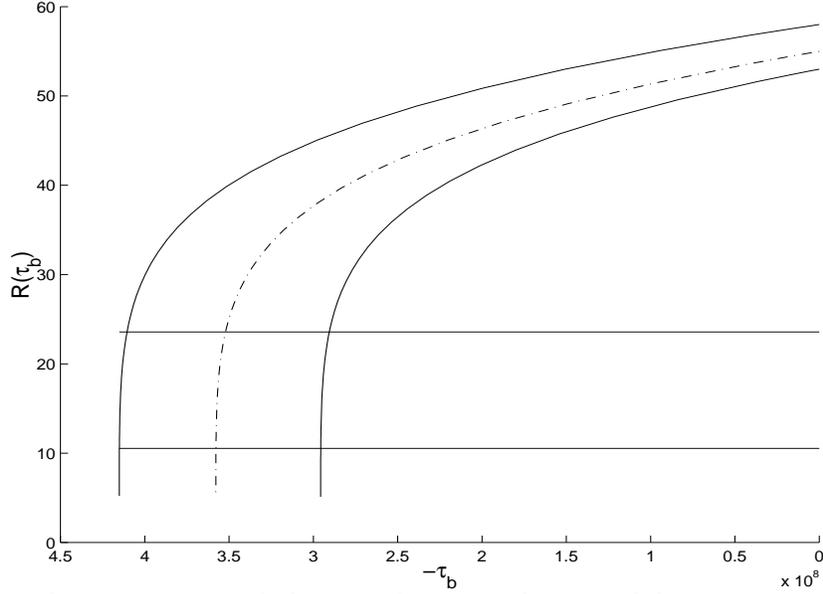}
\caption{\emph{The position of our various extend objects is shown as
a function of the proper time on each of them. The two solid curves
are the orbifold fixed planes while the broken line is a 5-brane. The
only physical part of the space is of course contained between the two
solid curves. We have chosen $\a_1 = -20$, $\a_2 = 10$, $\a_3 = 10$
for the brane charges and $M_1 = 0.04$, $M_2 = 0.05$ for the bulk mass
parameters. The two horizontal lines correspond to horizons in the
maximally extended versions of each of the two pieces of bulk. The
solution corresponds to a raised four-dimensional configuration when
all the branes are far above these lines.}}
\label{fig2}
\end{figure}

\begin{figure}[ht]\centering
\includegraphics[height=8cm,width=11cm, angle=0]{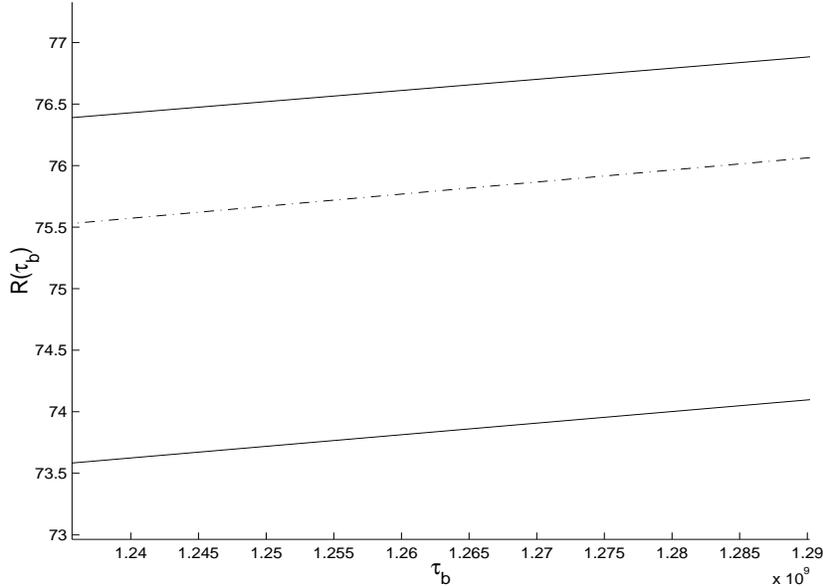}
\caption{\emph{The positions of the extended objects are shown for the
same choices of constants as in figure~\ref{fig2} but over different
time ranges. Clearly, the branes are at almost constant separation for
very long periods of time in this portion of the solution. This is due
to our choice of $\left[ M \right]$ as small. Throughout this time,
the 5-brane is roughly static with respect to the orbifold and we are
well above the horizons of the extended bulk metrics. Hence an
effective four-dimensional description is valid to good approximation
here.}}
\label{fig3}
\end{figure}

\begin{figure}[ht]\centering
\includegraphics[height=8cm,width=11cm, angle=0]{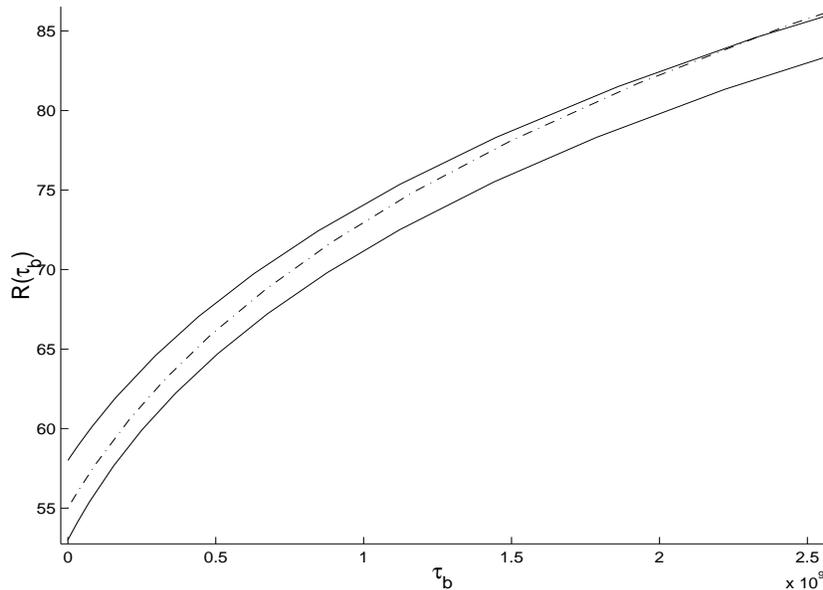}
\caption{\emph{The positions of the extended objects are shown for the
same choices of constants as in figures~\ref{fig2}-\ref{fig3} but
again over a different time ranges. It can be seen that the 5-brane
starts to move appreciably and eventually hits the uppermost fixed
point. Once the 5-brane starts to move relative to the fixed points in
this manner the four-dimensional description of the system breaks
down.}}
\label{fig4}
\end{figure}

Presented in figures~\ref{fig2}-\ref{fig4} are numerical solutions for
the brane motion as determined by equation~\eqref{breom}. The plots
are given in terms of $R$ and $\t_b$. These variables are the scale
factor and comoving time of the induced metric on the brane in
question and, it should be noted, are not the same as the
four-dimensional comoving time and scale factor.

The plots are all for the same values of parameters and integration
constants (as described in the caption of figure~\ref{fig2}) but are
plotted over different ranges. Figure~\ref{fig2} shows the overall
shape of the solutions. Figure~\ref{fig3} shows that for very large
changes in the time co-ordinates in the four-dimensional region the
branes barely move with respect to one another at all. This directly
leads us to the same conclusion in four-dimensional
units. Figure~\ref{fig4} shows the five-brane finally starting to move
appreciably relative to the orbifold fixed points. This results in
the brane impacting upon the upper fixed point as we saw was inevitable in
Section~\ref{solution}. The four-dimensional description is valid when
all the extended  objects are far above the horizon and before
the five-brane has begun to move appreciably relative to the orbifold
fixed points. In other words the four dimensional regime is roughly
the part of the motion shown in figure~\ref{fig3}. 
\vspace{2cm}

\vspace{0.4cm}

\vspace{0.4cm}

\vskip 0.4cm


\section{Causal Structure}

\begin{figure}[ht]\centering
\includegraphics[height=6cm,width=5cm]{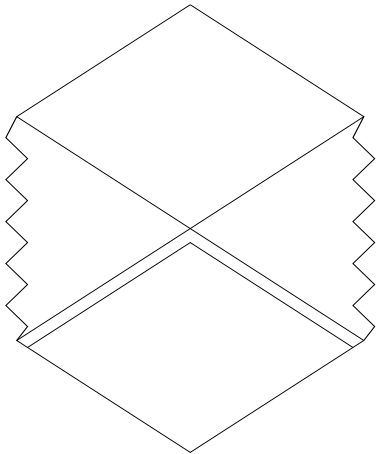}
\caption{\emph{Penrose diagram of maximally extended bulk segment.}}
\label{fig5}
\end{figure}

The Penrose diagram detailing the causal structure of a maximally
extended piece of the bulk spacetime we are employing is given in
figure~\ref{fig5}. The reader may be concerned that the horizon which
is drawn with a double line is of the same form as the
Reissner-Nordstr\"{o}m Cauchy horizon and as such is unstable in the
same sense. In fact we are not interested in a maximally extended
piece of the bulk but in two segments of such solutions sandwiched
between our various extended objects. This changes the causal
structure in such a way as to remove the unstable horizon.

\begin{figure}[ht]\centering
\includegraphics[height=6cm,width=5cm]{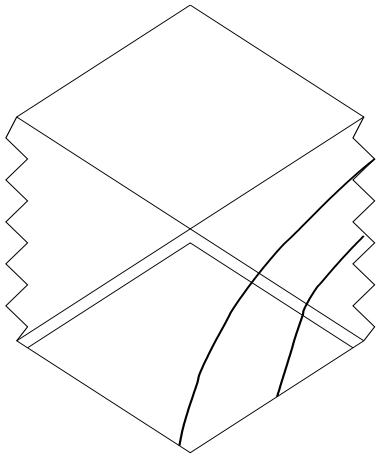}
\caption{\emph{Penrose diagram as in figure~\ref{fig5} with orbifold
fixed point trajectories included.}}
\label{fig6}
\end{figure}

This point is best illustrated by looking at the case without a
5-brane which is illustrated in figure~\ref{fig6}. Here the two new
lines represent the world volumes of the orbifold fixed planes. In the
solutions in this paper, the only physical space is that piece of the
bulk which lies between these two curves - the rest is discarded. This
of course has a profound effect on the causal structure of the
spacetime. The horizon which passes through the two fixed points was
only a horizon while we kept the whole spacetime. Specifically, when
we orbifold the space between the two boundary branes, we discard
portions of the singularities and hence the horizons cease to have
physical meaning. In fact, from the point of view of the physical
spacetime we are left with, the lines at $45^0$ on figure~\ref{fig6}
are no longer really horizons at all and so our fixed points do not
pass through a potentially singular unstable horizon.

Another way of seeing that the `horizon' is no longer unstable is that
the `paths of infinite blue shift' which cause the instability in the
Reissner-Nordstr\"{o}m case have been removed from the space time by
the cutting procedure.  Similar comments of course apply to cases where
5-branes are present. The global structure of the bulk solutions
without 5-branes was first discussed in~\cite{Chamblin:1999ea}.

\vskip 0.4cm

\vskip 0.4cm


\end{document}